\documentclass[11pt,letterpaper]{article}

\usepackage[T1]{fontenc}
\usepackage[english]{babel}
\usepackage{times}

\usepackage[letterpaper,margin=1in]{geometry}

\usepackage{amsmath,amstext,amsbsy,amsopn,amsthm,amsfonts,amssymb}
\usepackage{booktabs,array,colortbl}
\usepackage[]{graphicx}
\usepackage{rotate}
\usepackage{wrapfig}
\usepackage{caption}
\usepackage{subfigure}
\usepackage{lastpage}

\usepackage[small]{titlesec}
\usepackage{paralist}

\usepackage{verbatim} 
\usepackage[usenames,dvipsnames,svgnames]{xcolor}
\usepackage{tikz} 
\usepackage{pgfgantt}
\usepackage{pgfcalendar}
\usetikzlibrary{shadows}
\usetikzlibrary{shadings}
\usepackage{multirow}
\usepackage{algorithmic}
\usepackage{algorithm}

\usepackage{fancyhdr}
\providecommand{\keywords}[1]{\textbf{\textit{Keywords---}} #1}

\renewenvironment{thebibliography}[1]{%
	\begin{oldthebibliography}{#1}%
		\setlength{\parskip}{0.0cm}%
		\setlength{\itemsep}{0.0cm}%
	}%
	{%
	\end{oldthebibliography}%
}

\begin{document}

\title{\textbf{iVAMS 1.0}: Polynomial-Metamodel-Integrated Intelligent Verilog-AMS for Fast, Accurate Mixed-Signal Design Optimization}




\maketitle

\author{
	\begin{center}
		\begin{tabular}{cc}
			Saraju P. Mohanty & Elias Kougianos \\
			Computer Science and Engineering & Electrical Engineering Technology \\
			University of North Texas, Denton, TX 76203. & University of North Texas, Denton, TX 76203. \\
			Email: \texttt{saraju.mohanty@unt.edu} & Email: \texttt{elias.kougianos@unt.edu}
		\end{tabular}
	\end{center}
}

\cfoot{Page -- \thepage-of-\pageref{LastPage}}


\begin{abstract}
Electronic circuit behavioral models built with hardware description/modeling languages such as Verilog-AMS for system-level simulations are typically functional models. They do not capture the physical design (layout) information of the target design. Thus, their applications are limited to early design feasibility studies and/or pre-layout functional verification. Numerous iterations of post-layout design adjustments are usually required to ensure that design specifications are met with the presence of layout parasitics. In this paper a \textbf{paradigm shift of the current trend} is presented that integrates layout-level information (with full parasitics) in Verilog-AMS through metamodels such that system-level simulation of a mixed-signal circuit/system is realistic and as accurate as true parasitic netlist simulation. The simulations performed with these parasitic-aware models can be used to estimate system performance without layout iterations. We call this new form of Verilog-AMS as iVAMS (i.e. \textbf{Intelligent Verilog-AMS}). We call this iVAMS 1.0 as it is simple polynomial-metamodel integrated Intelligent Verilog-AMS. As a specific case study, a voltage-controlled oscillator (VCO) Verilog-AMS behavioral model and design flow are proposed to assist fast PLL design space exploration. The PLL simulation employing quadratic metamodels achieves approximately $10\times$ speedup compared to that employing the layout extracted, parasitic netlist. The simulations using this behavioral model attain high accuracy. The observed error for the simulated lock time and average power dissipation are 0.7~\% and 3~\%, respectively. This behavioral metamodel approach bridges the gap between layout-accurate but fast simulation and design space exploration. The proposed method also allows much shorter design verification and optimization to meet stringent time-to-market requirements. In the PLL optimization case study, 46~\% PLL power reduction was achieved using a differential evolution algorithm and the proposed layout-accurate behavioral model. Compared to the optimization using the layout netlist, the runtime using the behavioral model is reduced by 88.9~\%.
To the best of the authors' knowledge this is the first approach that brings layout-level accuracy to system-level design exploration and is applied towards mixed-signal design exploration.
\end{abstract}

\keywords{Metamodels, Surrogate Modeling, Mixed-Signal Design, Behavioral Simulation, Verilog-AMS Modeling, Intelligent Verilog-AMS, PLL, Design Exploration}

\section{Introduction}

\begin{figure}[htbp]
	\centering
	\includegraphics[width=0.69\textwidth]{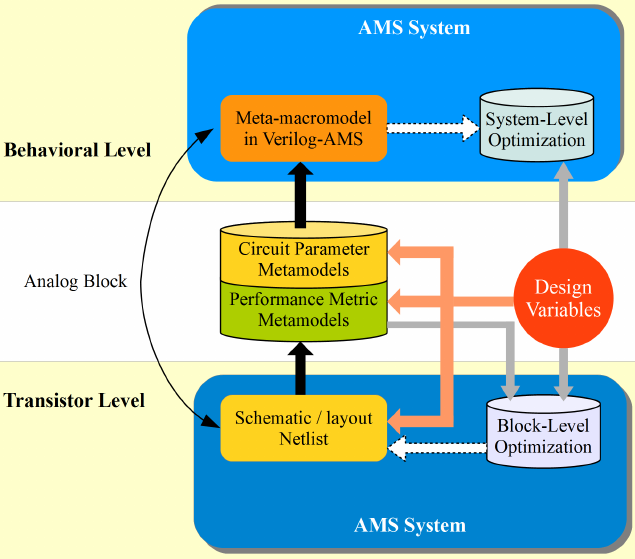}
	\caption{The concept of the proposed iVAMS.}
	\label{FIG:iVAMS_Idea}
\end{figure}

System-level modeling (using Verilog-AMS or VHDL-AMS) does not capture the physical design (layout) information of the target design as it is meant for \emph{fast behavioral simulation} only \cite{Mohanty_Book_2015_Mixed-Signal,Gielen_DATE_2019}. In particular, the effects of parasitics and process variation for nanoscale technology aggravate the situation. On the other hand, accurate simulations of nanoscale systems at the layout-level \emph{with full-blown parasitics (RCLK) are very slow} and even intractable for large systems \cite{Mohanty_Book_2015_Mixed-Signal,mohanty2015methodology}. As a \textbf{paradigm shift of existing trends}, we propose to incorporate layout-level information in Verilog-AMS through metamodels such that system-level simulation of a mixed-signal system is realistic and accurate in feasible time. The idea is presented in Fig. \ref{FIG:iVAMS_Idea}. The resulting layout-\textbf{I}ntelligent \textbf{V}erilog-\textbf{AMS} is called \textbf{iVAMS} \cite{Mohanty_Book_2015_Mixed-Signal,Zheng_DAC_2013,Zheng_ASAP_2013}. In this article, we will call this new approach iVAMS 1.0 as it is a simple polynomial-metamodel integrated Intelligent Verilog-AMS. Metamodels are models in the form of mathematical functions or algorithms which are generated from actual circuits (i.e., netlists) and are different from ``macromodels'' (which are simplified circuit models). iVAMS can completely decouple the design simulation flow to non-EDA tools, thus making the design-process \emph{very fast (with 10,000$\times$ speedup)} compared to the use of analog simulators \cite{mohanty2015intelligent,Mohanty_Invited-Talk_2012}.

Parasitics greatly degrade the performance of nano-CMOS circuit designs. They cause significant mismatch between schematic and layout circuit simulations. To account for the parasitic effects and achieve design closure, numerous iterations at the layout stage are usually required. Layout-accurate verification is the major obstacle because the iteration time is mainly spent on layout modification and simulation. Behavioral models that are capable of representing circuit layout have the potential to dramatically shorten the design cycle \cite{Mohanty_Book_2015_Mixed-Signal,GaritselovTSM2012,pam_2010,hun_2006,hsi_2006}. Techniques such as model order reduction \cite{woo_2004} were proposed to reduce the complexity of circuit models. Parasitic effects, however, are not discussed in most works due to the inherent inability of macromodel-derived behavioral models to account for them. Also, circuit models in these works are commonly implemented as Verilog-A modules rather than Verilog-AMS modules which are more efficient when used in conjunction with a fast digital simulator. Modeling techniques that do not account for both model compactness and layout-level accuracy can only have limited success. In this paper, we address both using a novel metamodeling based methodology.
An ultra-fast AMS design space exploration method based on  layout-accurate Verilog-AMS metamodels is proposed. It may be noted that \textbf{the terms macromodel and metamodel are often used interchangeably} in the literature. However, while macromodels are simplified models of a circuit and system that use the same simulator \cite{BasuVLSID2009}, metamodels are mathematical algorithms that can decouple the design and simulations to a pure behavioral tool such as MATLAB \cite{GaritselovTSM2012}.

The iVAMS based methodology proposed in this article aims at providing a way for the following:
\begin{enumerate}
	\item 
Capture post-layout parasitic effects; 
\item
Establish relationships between design variables and post-layout circuit behavior; and 
\item
Build parasitic-aware behavioral models for system-level simulation.
\end{enumerate}  
A major application of the proposed method is to complement the traditional top-down design flow. With the proposed method, the designer can examine the system-level impacts of design variables with layout parasitic effects, make adjustments to design variables, and re-simulate without re-doing the layout. Thus it greatly reduces the iterations from first layout design to the final one. The proposed method is based on metamodeling techniques.

A metamodeling technique has been explored for nano-CMOS AMS circuit design exploration \cite{GaritselovTSM2012}. The models built with this method accurately reflect parasitic effects. In the present work, the metamodeling approach is used to construct layout-accurate circuit behavioral models. These models can then be used for accurate and fast high-level simulation. To demonstrate this methodology, a VCO behavioral model is proposed based on this approach. This model is implemented using the Verilog-AMS language which enables fast simulations of a phase-locked loop (PLL). Combining metamodeling techniques and Verilog-AMS simulation, the design verification process achieves a large speedup and maintains reasonably high accuracy. Not only can the proposed Verilog-AMS behavioral model assist design verification of complex System-on-Chip (SoC) designs, but it also leverages design space exploration and optimization. A PLL design with an LC-tank VCO using 180 nm CMOS process is used to demonstrate the modeling technique, design flow, and optimization.
To demonstrate that the proposed method is compatible with state-of-the-art optimization algorithms, such as evolutionary algorithms (EA), we demonstrate PLL optimization using the proposed layout-accurate behavioral model with the powerful differential evolution (DE) algorithm.

The rest of this paper is organized as follows: Section~\ref{Sec:Contributions} describes the key ideas and contributions of this work. Section~\ref{Sec:Prior_Research} discusses previous works relevant to the techniques for accelerating AMS design simulation and design space exploration. Section~\ref{Sec:CPPLL_Metamodeling} presents the metamodeling technique and the proposed Verilog-AMS behavioral model. Section~\ref{Sec:PLL_Verilog-AMS-PAM} presents the PLL simulation flow and methodology with the proposed layout-accurate behavioral model. Section~\ref{Sec:PLL_Optimization} demonstrates the PLL optimization with the proposed behavioral model. Section~\ref{Sec:Conclusions} concludes this paper and discusses directions for future research.

\section{Contributions of this Paper}
\label{Sec:Contributions}

\begin{figure}[htbp]
\centering
\includegraphics[width=0.70\textwidth]{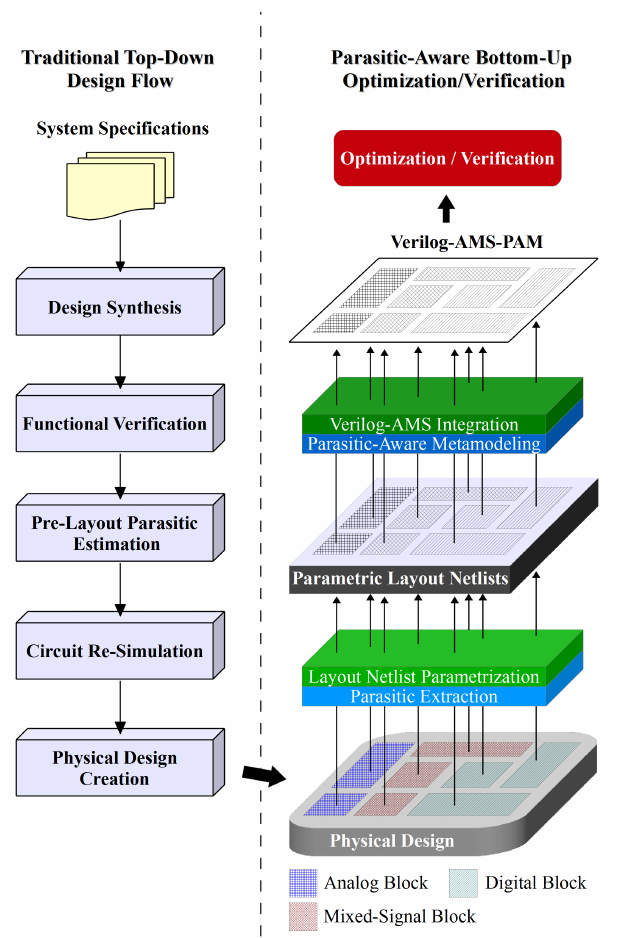}
\caption{The traditional top-down design flow and the proposed Verilog-AMS-PAM based parasitic-aware bottom-up optimization/verification flow.}
\label{fig:Verilog-AMS-PAM_Key-Idea}
\end{figure}

In this paper a \textbf{paradigm shift of the current trend} is presented that integrates physical design information (with full parasitics) in Verilog-AMS through parasitic-aware metamodels such that system-level simulation of a mixed-signal circuit/system is realistic and almost as accurate as the true parasitic netlist simulation. The Verilog-AMS module encapsulating the parasitic-aware metamodels is named Verilog-AMS-PAM. Verilog-AMS-PAM is an example of a specific instance of Intelligent Verilog-AMS (iVAMS) (to be considered as \textbf{iVAMS 1.0} in this article to suggest simple polynomial metamodel integration in Verilog-AMS) that bridged the gap between fast-inaccurate system-level simulation, and slow-accurate circuit-level or layout simulation. The key idea is depicted in Fig. \ref{fig:Verilog-AMS-PAM_Key-Idea}.

The AMS system design space exploration flow based on Verilog-AMS-PAM is essentially a parasitic-aware bottom-up approach complementary to the traditional top-down design flow. Parasitic awareness is introduced into the top-down flow through pre-layout parasitic estimation. The estimated parasitics are then inserted back to the circuit models and re-simulation is performed to adjust the design. Verilog-AMS-PAM provides an efficient way to perform accurate parasitic-aware AMS system optimization and verification.

The parasitic-aware bottom-up flow starts with extracting the parasitics from an initial physical design created at the end of the top-down flow. The parasitic-included netlist of each building block is then parameterized with respect to design and process variables. Metamodeling samples the response surfaces of the block to be modeled and generates mathematical models to approximate the entire response surfaces \cite{Mohanty_Book_2015_Mixed-Signal,mohanty2015intelligent}. Typically each block can be described by a set of metamodels. These metamodels are integrated into a Verilog-AMS module to behaviorally represent the block. Such a Verilog-AMS module is called Verilog-AMS-PAM. In the final step of the bottom-up flow, the Verilog-AMS-PAMs created for each block are assembled to form a system model. The system model created this way is compact and automatically includes the full parasitic circuit response. Thus it can be used for fast and accurate system-level optimization and verification. Once the optimized design parameters have been obtained, the layout is adjusted to these new geometric dimensions (for a sizing problem). The introduced perturbation is generally small and the resized layout is very near the optimal target. This has been verified in this work and our previous related research \cite{GhaiTVLSI2009Sep}. 

The {\bf novel contributions} of this article are the following:
\begin{enumerate}

\item
The concept of Verilog-AMS-PAM (or iVAMS 1.0) is introduced to facilitate efficient mixed-signal design space exploration.

\item
An effective flow to create compact and layout-accurate circuit behavioral models for system optimization and verification is proposed coupled with
a fast AMS design simulation approach that is compatible with state-of-the-art optimization algorithms.

\item
An accurate and efficient quadratic polynomial metamodel for a 180 nm LC-VCO design is developed.
Implementation details of the Verilog-AMS-PAM creation for the VCO and the behavioral model of a PLL are also presented.

\item
The accuracy and speed of the proposed Verilog-AMS-PAM based AMS design simulation is discussed through the PLL case study.

\item
Metamodel-integrated PLL simulations are presented and the accuracy and speed of the proposed VCO behavioral Verilog-AMS model are discussed.

\item
An optimization flow within the iVAMS 1.0 framework is demonstrated  with a PLL as a case study.

\end{enumerate}

\section{Related Prior Research}
\label{Sec:Prior_Research}

Design space exploration employing traditional SPICE simulation \cite{phe_2000} relies on efficient optimization algorithms to reduce the number of iterations. Various techniques had been developed to speed up these simulations \cite{lor_2011}. Fast-SPICE simulators \cite{gul_2009,rew_2011} offer a speedup of roughly one order of magnitude, which is often insufficient. Another class of approach is to construct macromodels and/or metamodels to represent AMS designs and to perform design space exploration over these models. It is important for the models to support the inclusion of parasitic effects \cite{buh_2006}. One such model is the parasitic-aware symbolic model proposed in \cite{ran_2004}. However, this symbolic model is limited to modeling small-signal behavior and is inefficient when the circuit sizes are large.  In \cite{li_2003}, a technique employing Volterra series based models together with model order reduction and pruning was proposed, but it is restricted to weakly nonlinear circuits.

In the literature, macromodels have been referred to as white-box models that carry a certain amount of physical information of the design, and as black-box models that approximate the circuit behavior. They are typically either simplified structures of the design \cite{wei_2005} or analytical equations derived based on the designers' knowledge of the circuit \cite{koh_1990,sre_2008}. Metamodels, also called surrogate models, are gaining increasing attention in circuit design and are black-box models. Metamodeling attempts to approximate the response surfaces of the design by first sampling the design space and then fitting a chosen metamodel to the sampled point responses. The metamodels that have been used for AMS designs include polynomials \cite{dae_2003,li_2007}, splines \cite{vem_2004}, support vector machines \cite{deb_2003}, artificial neural networks \cite{wol_2003}, and Kriging \cite{yu_2007,oko_2012}. It is worth mentioning that historically the response surface methodology (RSM) in the literature typically refers to metamodeling with low-order polynomials (quadratic at most). While low-order polynomials are limited to a small number of variables and small design spaces, state-of-the-art metamodeling techniques employing intelligent models can handle large and highly nonlinear design spaces. Despite the success of the aforementioned metamodels for block-level sizing and optimization, they were not intended for behavioral model construction to support higher-level system verification or design space exploration.

Efforts such as \cite{liu2002remembrance, li_2003, gu2010generalized, yel_2012} have been made to advance the modeling techniques. In this work, we focus on the techniques that consider parasitic effects. Behavioral models and their construction using hardware description languages (HDLs) such as SystemC-AMS, VHDL-AMS, and Verilog-AMS for efficient AMS system design space exploration have become popular. The VHDL-AMS op-amp models presented in \cite{jan_2006} and \cite{sab_2007} take into account nonidealities such as parasitics. In \cite{jan_2006}, concepts of exploring analog design spaces with parasitic-included behavioral models were discussed. The model in \cite{sab_2007} is valid with various loads and accounts for output nonlinear behavior. The limitation is that it requires device information such as MOSFET operation region. Thus it is difficult to apply this technique to complex nanometer circuits. A semi-symbolic analysis technique using affine arithmetic was proposed in \cite{gri_2005} to model and analyze AMS system performance degradation caused by nonidealities. The system was described in SystemC-AMS. This technique has the same accuracy limitation as macromodels. Also, SystemC-AMS suffers from the difficulty of modeling nonideal and nonlinear systems, and thus is less accurate compared to VHDL-AMS and Verilog-AMS.
For this reason, mixed SystemC-AMS, VHDL-AMS, and Verilog-AMS modeling was suggested in \cite{zai_2009}. In this work, we adopted Verilog-AMS and metamodeling to construct a compact and layout-accurate VCO behavioral model used for PLL design space exploration.

Many PLL and VCO behavioral models exist in literature. Still, most of them were either not intended for layout-accuracy or are too simple to capture the nonlinearity of mixed-signal design spaces. Verilog-A behavioral modules of linear VCOs were used in \cite{jin_2010} for PLL jitter characterization and in \cite{zou_2005} for aiding a hierarchical CPPLL sizing method. No parasitic effects were included in these models. A characterization technique is developed in \cite{kuo_2006} to extract circuit parameters, including parasitic effects. The authors also adopted the linear VCO model which may be sufficient for performing verification on fixed designs, but is overoptimistic for design exploration since the VCO linearity condition is not always valid. The VCO behavioral models developed in \cite{ali_2009,har_2010} used lookup-tables (LUTs) inside Verilog-A modules. LUTs only hold a limited number of simulated sample points. The circuit responses in-between these points are estimated using interpolation. Interpolation limits the accuracy of LUT models while metamodeling offers higher accuracy overall. An event-driven analog modeling approach was proposed in \cite{wan_2009} which used the Verilog-AMS {\tt{wreal}} data type to improve the model efficiency. However, it is not clear how the VCO gain and output frequency were modeled.

\section{Proposed Methodology for iVAMS 1.0 (i.e. Verilog-AMS-PAM) Generation}
\label{Sec:CPPLL_Metamodeling}

\subsection{Proposed Methodology for iVAMS 1.0 (i.e. Verilog-AMS-PAM)}

\begin{figure}[htbp]
\centering
\includegraphics[width=0.80\textwidth]{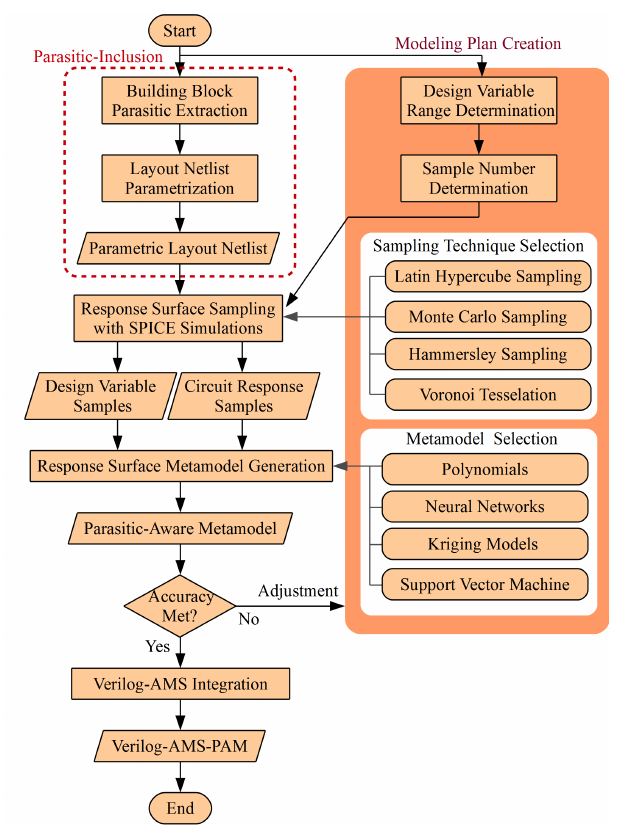}
\caption{Verilog-AMS-PAM Generation Flow.}
\label{fig:Verilog-AMS-PAM_Generation}
\end{figure}

The generation of Verilog-AMS-PAM for a circuit block involves parametric layout netlist creation, modeling plan creation, circuit response surface sampling, response surface metamodel generation, and Verilog-AMS integration.
The Verilog-AMS-PAM generation flow is depicted in Fig.~\ref{fig:Verilog-AMS-PAM_Generation}. Parametric layout netlist generation is the key step to include the parasitic effects. It also ensures that the model estimating the circuit response surfaces is the closest to the silicon results. The modeling plan creation includes determination of the design variable ranges, selecting a sampling technique and a metamodel suitable to model the particular circuit block. The Latin Hypercube Sampling (LHS) technique ensures that the samples are distributed over the entire design space and that each sample is distributed randomly in the pre-defined sub-space. It achieves a good trade-off between uniform and random sampling and thus is widely used.

Metamodel selection aims at selecting a metamodel that can meet the accuracy requirements with minimum complexity. This process can be aided by prior knowledge about the circuit block. For example, in our case study, quadratic polynomial metamodels are used to model an LC VCO. For more complex analog blocks with high nonlinearity, intelligent metamodels such as artificial neural networks \cite{zheng_2012_isvlsi} can be used. For circuit blocks without prior knowledge, coarse sampling can be performed to facilitate a quick complexity analysis. Alternatively, metamodels not restricted to a specific type can be created via canonical-form functions as in \cite{mcc_2009} where the model selection was formulated as a multi-objective optimization problem. The model generation settings shown in Fig.~\ref{fig:Verilog-AMS-PAM_Generation} can be adjusted easily to select and compare different kinds of models. 
An optimization algorithm was employed to find the metamodel Pareto front that minimizes the model complexity and the prediction error. The cost is the increased modeling effort.

The response surface sampling is best performed using SPICE simulations and the parasitic-included layout netlist to ensure high accuracy. The selected metamodel is fitted to the sampled data by tuning the model coefficients. If the accuracy is not satisfied, adjustment can be made to the modeling plan such as increasing the sample size, changing the metamodel architecture, or using a different metamodel. Since the layout-level information is included in the parametric netlist, the resultant circuit block metamodels are layout-accurate parasitic-aware metamodels. These metamodels are described in Verilog-AMS and embedded in the Verilog-AMS module to construct the behavioral model for the block.

\subsection{High-level Description and Modeling of Mixed-Signal Design - A PLL Case Study}

\begin{figure}[htbp]
\centering
\includegraphics[width=0.60\textwidth]{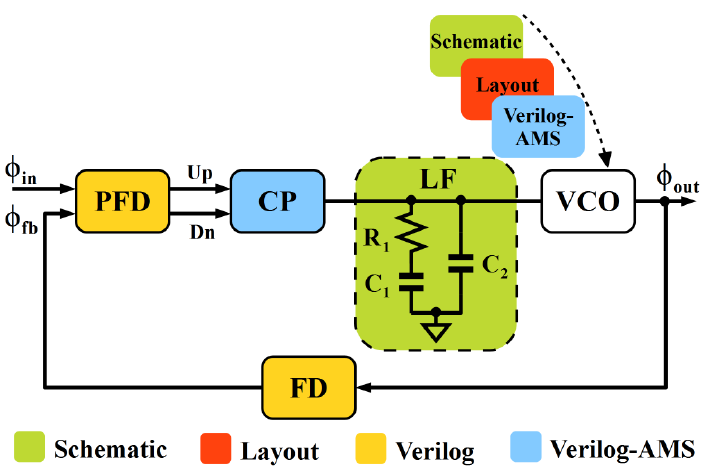}
\caption{Block level representation of CPPLL configuration.}
\label{fig:PLL_Block_Diagram}
\end{figure}

A typical charge pump PLL (CPPLL) consists of a phase/frequency detector (PFD), a charge-pump (CP), a loop filter (LF), and a VCO. If the PLL needs to perform frequency synthesis, a frequency divider (FD) will also be employed. The system level topology of a CPPLL is shown in Fig.~\ref{fig:PLL_Block_Diagram}. A CPPLL is a mixed-signal system. The CP, LF, and VCO directly deal with the analog signal therefore are the most critical parts. A more comprehensive PLL analysis can be found in \cite{book_pll_raz1996}. This case study focuses on developing a VCO behavioral model that can accurately represent the VCO \emph{physical} design. The model is constructed using the Verilog-AMS language to enable fast design exploration. The other parts of the PLL are modeled with HDLs or at the schematic level in order to simulate the whole PLL system.

Fig.~\ref{fig:PLL_Block_Diagram} illustrates the CPPLL configuration in this paper. The frequency of the FD output $\phi_{fb}$ is $1/N$ of that of the VCO output $\phi_{out}$, where $N$ is the FD division ratio. The PFD activates its output $Up$ or $Dn$ to vary the VCO output until $\phi_{fb}$ and $\phi_{in}$ are aligned and have the same frequency. They introduce nonidealities to the system via their signal delay, and the rise/fall time. These nonidealities can be easily described in the digital domain. Thus the behavior of these two blocks is implemented using the Verilog language. The CP has digital inputs and analog output so it is implemented as a Verilog-AMS module. Portions of the source code for the PFD and the CP are shown in Algorithm\ref{ALG:pfdcode}.

\begin{algorithm}[t]
\caption{Verilog-AMS model for the PFD and the CP.}
\label{ALG:pfdcode}
\centering
\begin{verbatim}
module PFD(up, dn, clkin, clkfb); //Verilog
... ...
   assign #1 reset = up && dn;
   always @(posedge clkin or posedge reset)
   begin
      if (reset) up <= 1'b0;
      else up <= 1'b1;
   end
   always @(posedge clkfb or posedge reset)
   begin
      if (reset) dn <= 1'b0;
      else dn <= 1'b1;
   end
endmodule //End PFD

module CP (out, up, dn); //Verilog-AMS
   parameter real cur = 50u; //output current
   ... ...
   analog begin
   @(initial_step) iout = 0.0;
   if (dn && !up && (V(out) > gnd))
      iout = -cur;
   else if (!dn && up && (V(out) < vdd))
      iout = cur;
   else iout = 0;
   I(out) <+ -transition(iout, 0.0, 2p, 2p);
   end
endmodule //End CP
\end{verbatim}
\end{algorithm}

\begin{figure}[t]
\centering
\subfigure[Schematic view]{\includegraphics[height=0.45\textwidth]{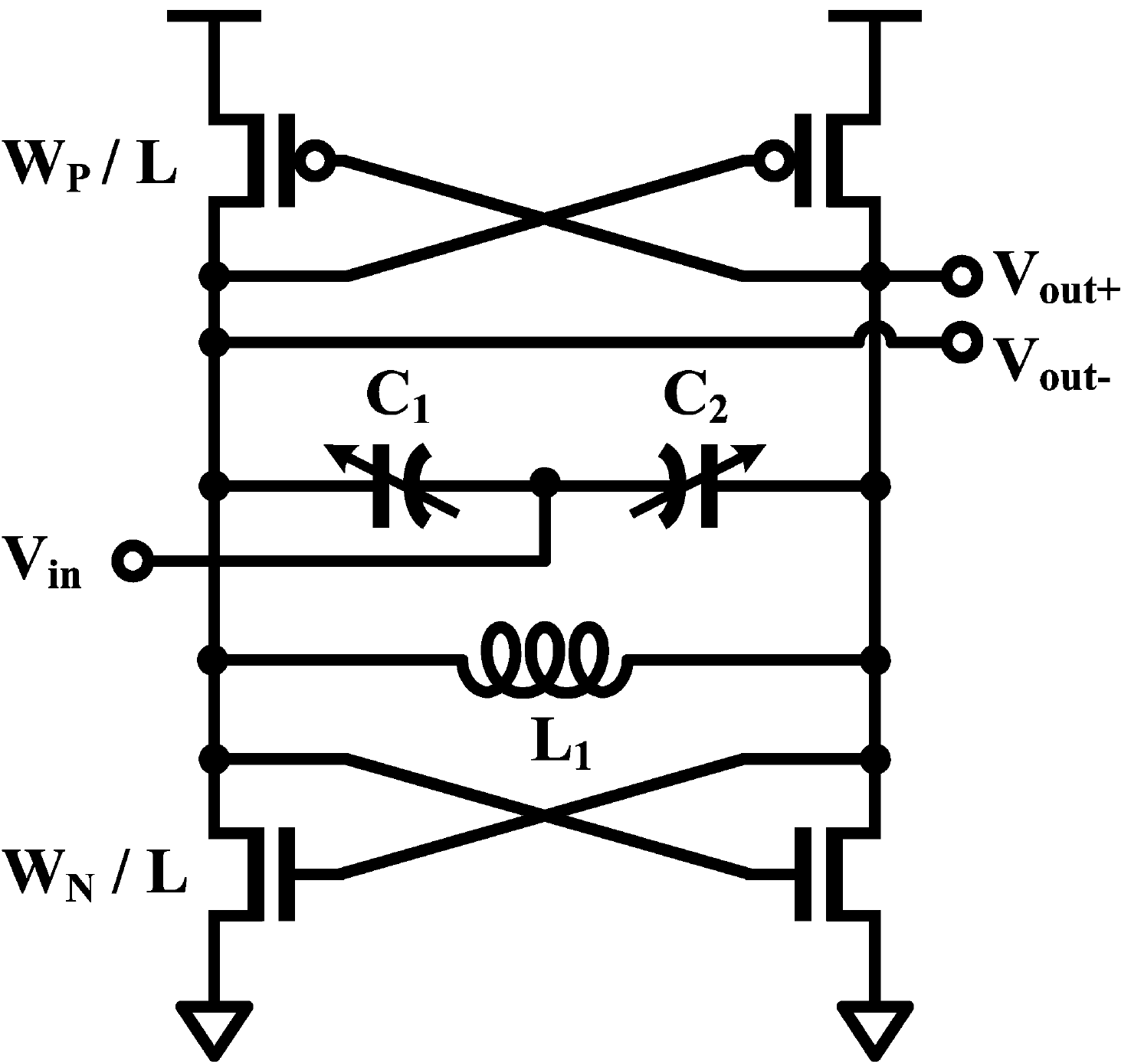}
\label{fig_vcosch}}
\subfigure[Layout view]{\includegraphics[height=0.45\textwidth]{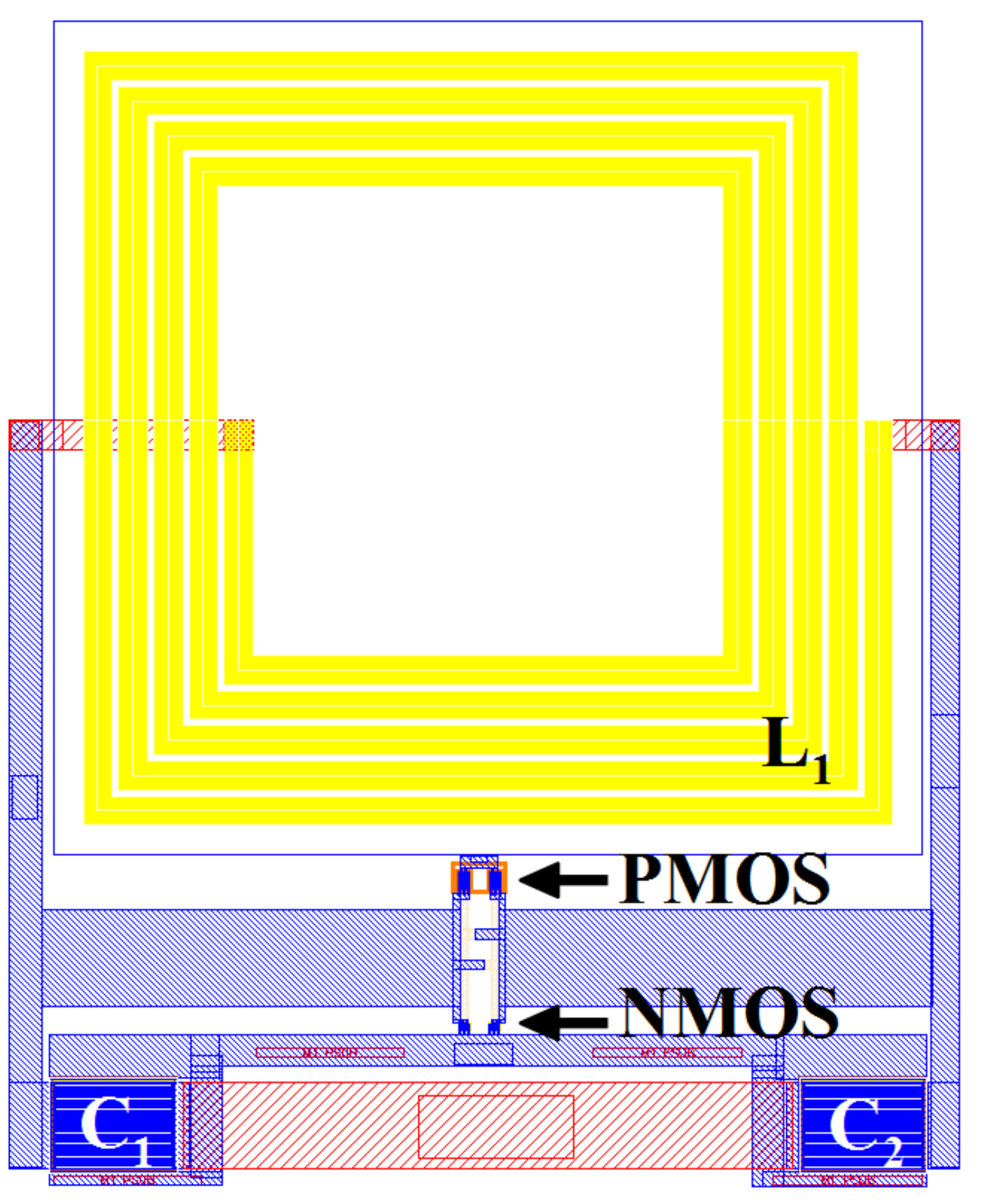}
\label{fig_vcolay}}
\caption{The LC VCO schematic and layout views. $L = 180$ nm; $W_P = 20$ $\mu$m; $W_N = 10$ $\mu$m.}
\label{fig:LC-VCO}
\end{figure}

Three different views have been implemented for the VCO: (1) schematic, (2) layout with parasitics, and (3) parasitic-aware Verilog-AMS (Verilog-AMS-PAM). Fig.~\ref{fig:LC-VCO} shows the schematic and layout views of the LC VCO design. Both schematic and layout views use SPICE models for simulation. While the layout view includes the parasitic elements and therefore takes longer to simulate, it results in a much better estimate of the real silicon performance. Table~\ref{TBL:LC-VCO} lists the number of elements in the schematic view and parasitic extracted layout view. The parasitics consist of Resistance (R), Capacitance (C), self inductance (L), and mutual inductance (K). Other blocks such as the FD and the CP can also be modeled using the Verilog-AMS-PAM technique, but in this work we only focus on the VCO to keep the case study simple yet effective.

\begin{table}[htbp]
\caption{Element Counts for The LC VCO Schematic and Layout Views.}
\label{TBL:LC-VCO}
\centering
\begin{tabular}{|c|c|c|}
\hline
{} & \bfseries  Schematic & \bfseries Layout \\
\hline \hline
 Transistor & 4 & 4 \\
\hline
 Inductor & 1 & 10 \\
\hline
 Capacitor & 2 & 38 \\
\hline
 Resistor & 0 & 560 \\
\hline
Total & 7 & 612 \\
\hline
\end{tabular}
\end{table}


\subsection{iVAMS 1.0 (or Verilog-AMS-PAM) for VCO}
\label{Subsec:VCO_Verilog-AMS-PAM}

The VCO behavior is mainly determined by its voltage frequency transfer curve. A common way to model a VCO is to assume that it is perfectly linear and model it with the following:
\begin{equation}
\label{eqn_vco}
f_{osc} = f_0 + K_{VCO} V_C,
\end{equation}
where $f_{osc}$ is the oscillation frequency, $f_0$ is the center frequency, $K_{VCO}$ is the gain, and $V_C$ is the control voltage at the VCO input. To account for non-linearities and layout parasitics, the metamodeling approach suggested in \cite{GaritselovTSM2012} is used. Polynomial metamodels are selected in this work for integration as they have the following advantages: (1) they are simple closed form equations which are easy to implement; (2) their form is flexible so that one can quickly examine and compare the accuracy of polynomial models with different degree; (3) they have been widely used and their properties are well understood. The polynomial metamodel used in this paper is as follows:
\begin{equation}
\label{eqn_poly}
f(\mathbf{x}) = \sum\limits_{i = 0}^{K - 1} \beta_i {x_1}^{p_{1i}} {x_2}^{p_{2i}} {x_3}^{p_{3i}},
\end{equation}
where $x_1$, $x_2$, and $x_3$ are three input variables corresponding to $W_P$, $W_N$, and $V_C$ in this work, respectively. $W_P$ and $W_N$ are the PMOS and NMOS widths, respectively. $K$ is the number of basis functions this model has and $\beta_i$ is the coefficient for the $i$-th basis function. $f(\mathbf{x})$ is the output that approximates the true model. In order to construct the metamodel for a given VCO design, for each basis function the coefficient $\beta_i$ and the power terms $p_{1i}$, $p_{2i}$, and $p_{3i}$ for each input variable need to be obtained. This is done in three steps: first, a set of input variables $[x_1$ $x_2$ $x_3]$ is generated using the Latin Hypercube Sampling (LHS) technique; second, circuit simulations are performed and the outputs for each set of inputs are saved; third, with the inputs and outputs from previous steps, the coefficients and the power terms that lead to a model with good fit are computed. In order to incorporate the parasitic effects into the model without repeating the layout for each simulation, the netlist for the extracted layout view is parameterized for $W_P$ and $W_N$. Algorithm~\ref{ALG:VCO_netlist} shows a portion of the parameterized layout netlist where the original PMOS model has been replaced with a parameterized one. The netlist is in the form of Verilog-AMS and thus can be accepted by the AMS simulator. As has been shown by our previous work in \cite{GhaiTVLSI2009Sep}, the parameterized layout netlist maintains high fidelity when compared to a real layout netlist. Thus parasitic re-extraction is not needed for new set of $W_P$ and $W_N$.

\begin{algorithm}[htbp]
\caption{A portion of the parameterized netlist for the VCO layout view.}
\label{ALG:VCO_netlist}
\centering
\begin{verbatim}
... ...
inductor #(.l(8.244e-11)) l1_291 (\291:RLJUNC_ ...
inductor #(.l(8.797e-11)) l1_2 (\2:RLJUNC_J1 , ...
pmos1 #(
.w(((cds_globals.Wp) / (4))), .l(cds_globals.L),
.as((((cds_globals.Wp)/(4))<599.5n) ? (((((200 ...
.ad((((cds_globals.Wp) / (4)) < 599.5n) ? ((fl ...
.ps((((cds_globals.Wp) / (4)) < 599.5n) ? (((...
.pd((((cds_globals.Wp) / (4)) < 599.5n) ? ((fl ...
... ..., .m("(1)*(4)"))
PM1 (\10:Voutp , \6:Voutn , cds_globals.\vdd! ,
cds_globals.\vdd! );
... ...
\end{verbatim}
\end{algorithm}

In this work, the VCO output frequency and its power consumption are of interest. Therefore two respective metamodels are constructed. They share the same power terms for the input variables, while the coefficients $\beta_i$ in the two models are different. After these values are computed, they are written into a text file which will be read by the VCO Verilog-AMS module to implement the model. A quadratic polynomial metamodel with first order interaction has been implemented. 
Table~\ref{TBL:Polynomial_Metamodel} shows the layout of the text file storing the values for the power terms and the coefficients for this model obtained from 100 samples. In the table, $\beta_{i,f}$ and $\beta_{i,p}$ are the coefficients for the frequency and power consumption models, respectively. These values are read into the Verilog-AMS module during the {\tt{initial}} block.

\begin{table}[htbp]
\caption{Layout of the text file storing the power terms and coefficients for the VCO quadratic polynomial metamodel}
\label{TBL:Polynomial_Metamodel}
\centering
\begin{tabular}{c|cccrr|}
\multicolumn{1}{r}{$i$} & $p_{1i}$ & $p_{2i}$ & $p_{3i}$ & \multicolumn{1}{c}{$\beta_{i,f}$} & \multicolumn{1}{c}{$\beta_{i,p}$} \\
\cline{2-6}
0 & 0, & 0, & 0, & 2.113e+009, & 1.385e-005 \\
1 & 1, & 0, & 0, & -3.214e+012, & 44.459e+000 \\
2 & 2, & 0, & 0, & 3.456e+016, & -2.804e+005 \\
3 & 0, & 1, & 0, & 6.869e+012, & 39.729e+000 \\
4 & 1, & 1, & 0, & -1.021e+017, & 2.911e+005 \\
5 & 0, & 2, & 0, & -2.071e+017, & -1.080e+006 \\
6 & 0, & 0, & 1, & 3.513e+008, & -8.271e-004 \\
7 & 1, & 0, & 1, & -2.565e+012, & -31.282e+000 \\
8 & 0, & 1, & 1, & -5.331e+012, & -11.392e+000 \\
9 & 0, & 0, & 2, & 0.000e+000, & 1.041e-003 \\
\cline{2-6}
\end{tabular}
\end{table}

Algorithm \ref{ALG:VCO_Verilog-AMS-PAM} shows a portion of the VCO Verilog-AMS module. The part of the basis function related to the input variables $W_P$ and $W_N$ is constructed in the \verb+initial+ block. The remainder of the basis functions are constructed in the \verb+always+ block since the third variable $V_C$ needs to be updated continuously during the simulation. The output signal of this module is implemented to be of digital logic type to reduce the computational cost. As in the PFD and FD modules, the non-idealities associated with this output signal can be modeled in the digital domain.

\begin{algorithm}[htbp]
\caption{Illustration of \textbf{iVAMS 1.0 } - Verilog-AMS integrated with Parasitic-Aware Metamodels (Verilog-AMS-PAM) for the LC-VCO.}
\label{ALG:VCO_Verilog-AMS-PAM}
\centering
\begin{verbatim}
`timescale 10ps / 1ps
`include "disciplines.vams"
module vco_metamodel (out, in);
... ...
parameter integer K;
initial
begin
    out = 0;   //Initialize vco digital output
    ... ... //Declare ports and data types
    metaf = $fopen("metamodel.csv", "r");
    while (!$feof(metaf))
    begin
      readfile = $fscanf(metaf,
                 "%e,%e,%e,%e,%e\n",
                 p1, p2, p3, betaf, betap);
      bf[i] = pow(wp,p1) * pow(wn,p2) * betaf;
      bp[i] = pow(wp,p1) * pow(wn,p2) * betap;
      pv[i] = p3;
      i = i + 1;
    end
    $fclose(metaf);
    ... ...
end
always
begin
    vc = V(in);
    ... ...
    freq = 0;
    power = 0;
    for (i = 1; i <= K; i = i + 1)
    begin
      freq = freq + bf[i] * pow(vc, pv[i]);
      power = power + bp[i] * pow(vc, pv[i]);
    end
    ... ...
    #(0.5 / freq / 10p)
    out = ~out;
end
... ...
endmodule
\end{verbatim}
\end{algorithm}

\begin{figure}[htbp]
\centering
\includegraphics[width=0.55\textwidth]{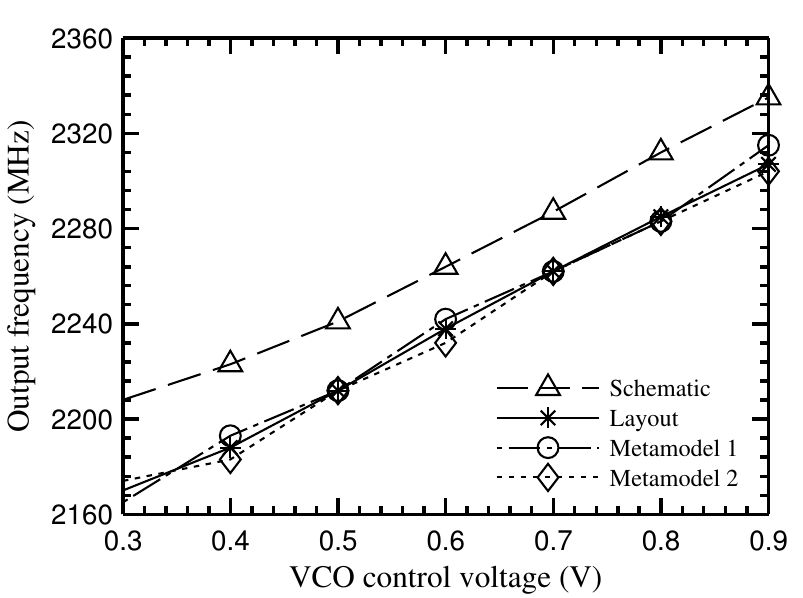}
\caption{VCO transfer curves for three different views.}
\label{fig:vcotf}
\end{figure}

This Verilog-AMS module can be easily reconfigured for metamodels with different degrees by changing the parameter \verb+K+. In Fig.~\ref{fig:vcotf}, the simulation results of the VCO transfer curves for the design in Fig.~\ref{fig:LC-VCO} are shown. The parasitics cause a large difference between the schematic and layout results both in the VCO center frequency and the gain. Metamodel 1 is the Verilog-AMS module with the quadratic model from 100 samples. Metamodel 2 is the module with a 5-th degree polynomial model from 500 samples. Metamodel 2 does not improve significantly over Metamodel 1. Thus Metamodel 1 is used in the PLL simulations shown in Sections~\ref{Sec:PLL_Verilog-AMS-PAM} and \ref{Sec:PLL_Optimization}. Differences between the transfer curves of layout and metamodel Verilog-AMS views can still be observed, which means a better metamodel may be used to further improve the accuracy. However, as will be seen in Section~\ref{Sec:PLL_Verilog-AMS-PAM}, this polynomial metamodel is sufficient for system level PLL verification to simulate lock time and average power dissipation.

\section{iVAMS 1.0 (i.e. Polynomal-Metamodel-Integrated Intelligent Verilog-AMS) based Simulation of PLL}
\label{Sec:PLL_Verilog-AMS-PAM}

In this section, we demonstrate PLL simulations with the VCO design shown in Fig.~\ref{fig:LC-VCO}. The simulation results for the PLL with Verilog-AMS-PAM and schematic and layout SPICE models are compared. The PLL structure shown in Fig.~\ref{fig:PLL_Block_Diagram} is used. The PFD and FD are in Verilog view. The CP is in Verilog-AMS view and the LF is in schematic view. The views for these blocks were not changed throughout the simulations. The VCO view was changed from schematic, to layout with parasitics, and then to Verilog-AMS views. Two Verilog-AMS views have been implemented--one for the linear model and one for the quadratic metamodel proposed in Section~\ref{Subsec:VCO_Verilog-AMS-PAM}. The results for different VCO views are obtained.

\begin{figure}[htbp]
\centering
\includegraphics[width=0.55\textwidth]{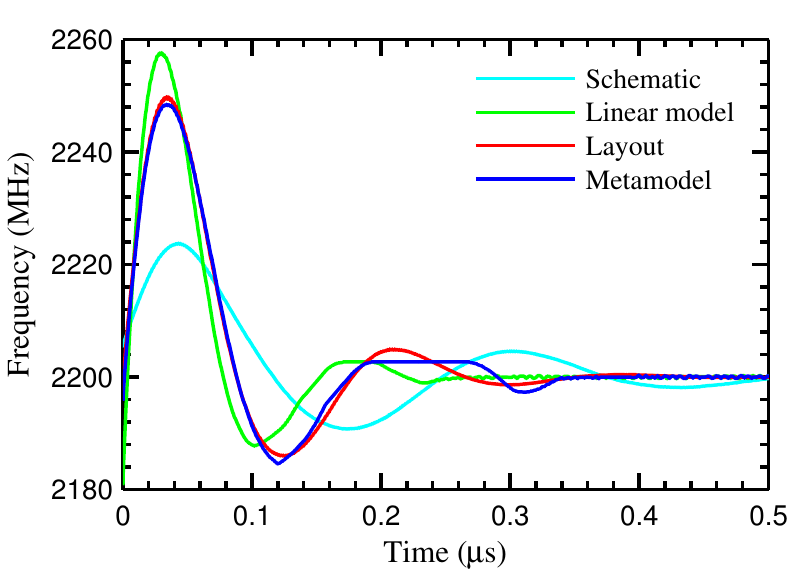}
\caption{PLL output frequency from AMS simulation with three different VCO views.}
\label{fig:frequency_plot}
\end{figure}

A 550 MHz input clock $\phi_{in}$ is assigned to the PLL input. The FD has a division ratio of 4. Thus the desired frequency for the PLL output clock $\phi_{out}$ is 2200 MHz. Fig.~\ref{fig:frequency_plot} shows the $\phi_{out}$ frequencies from 500 ns transient simulations with different VCO views. Although the PLLs with different VCO views are all able to lock to the same correct frequency, the one with the schematic view shows quite different locking behavior compared to the one with the layout view. This mismatch is due to the parasitic effects which greatly change the VCO transient characteristics. The one with the linear model shows improvements over the the schematic since the parasitics have been taken into account. However, it still has significant errors, for example, in the lock time. The PLL with the metamodel Verilog-AMS view offers the best approximation of the true model and accurately estimated the lock time. 
To further understand the behavior of the PLL with different VCO views, the critical analog signal $V_C$ was inspected.

\begin{figure}[t]
\centering
\includegraphics[width=0.55\textwidth]{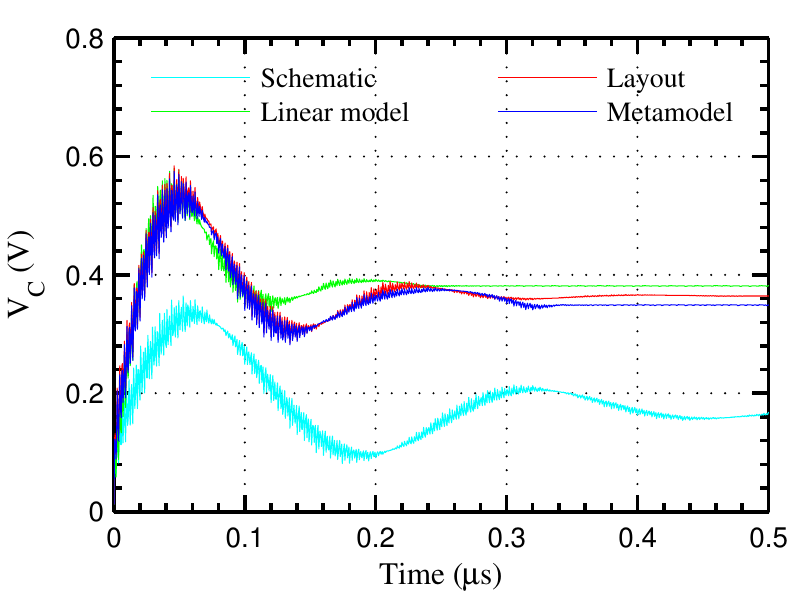}
\caption{VCO control voltage waveforms from PLL simulations.}
\label{fig:vcplot}
\end{figure}

Fig.~\ref{fig:vcplot} compares the $V_C$ waveform from the four simulations. Again, the metamodel Verilog-AMS view provides an excellent approximation of the layout view behavior. The PLL with the schematic VCO view can just barely lock to 2200 MHz since $V_C$ is approaching the NMOS threshold. This shows that the center frequency and the gain of the schematic VCO view are very different from the layout one. These further confirm the VCO transfer curves plotted in Fig.~\ref{fig:vcotf}.

\begin{figure}[htbp]
\centering
\includegraphics[width=0.55\textwidth]{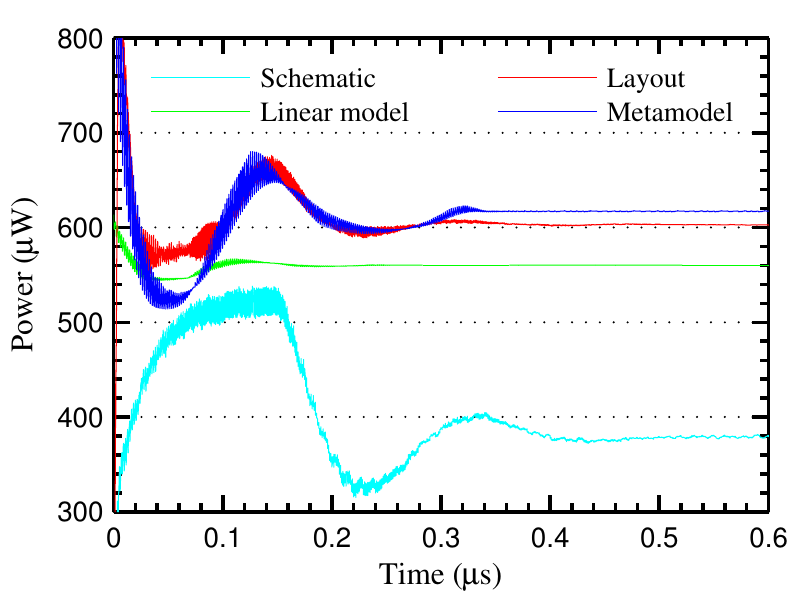}
\caption{Average VCO power consumption per 50 cycles.}
\label{fig:powplot}
\end{figure}

The Verilog-AMS metamodel also facilitates estimation of power consumption. Fig.~\ref{fig:powplot} shows the average VCO power consumption per fifty cycles in the four simulations. It once again confirms that the Verilog-AMS metamodel can model its layout counterpart very accurately compared to a linear model. Table~\ref{TBL:pllsim} summarizes the PLL simulation results and compares the accuracy of the linear model and the proposed metamodel.

In Table~\ref{TBL:pllsim}, the estimated PLL lock time is listed. The one from the simulation with the VCO layout view serves as the true model. The errors resulting from the other two models are computed. The metamodel achieves a very low error rate of $0.7~\%$, while the linear model causes a large error of $31.7~\%$.  $\mathbf{f_{Locked}}$ is the PLL output frequency when it is locked. $\mathbf{P_{Locked}}$ is the average VCO power consumption when the PLL is locked. Again, the metamodels give an accurate estimation of the power dissipation. The $\mathbf{V_C}$ root-mean-square error (RMSE) of the models for the 500 ns simulations are also listed.

\begin{table}[htbp]
\caption{Comparison of PLL Simulations with Different VCO Modules.}
\label{TBL:pllsim}
\centering
\begin{tabular}{|c|c|c|c|}
\hline
{} & \bfseries  Layout & \bfseries Linear Model & \bfseries Metamodel \\
\hline \hline
{Lock time} (ns) & 335.4 & 229.1 & 332.9 \\
 Error~\% & 0.0~\% & 31.7~\% & 0.7~\% \\
\hline
${f_{Locked}}$ (MHz) & 2199.99 & 2199.99 & 2199.99 \\
 Error~\% & 0.0~\% & 0.0~\% & 0.0~\% \\
\hline
${P_{Locked}}$ ($\mu$W) & 602 & 560 & 620 \\
 Error~\% & 0.0~\% & 7.0~\% & 3.0~\% \\
\hline
${V_C}$ RMSE (mV) & 0 & 33.508 & 10.889 \\
\hline
\end{tabular}
\end{table}

Table~\ref{TBL:speed} compares the runtimes for the PLL transient simulations. The Verilog-AMS metamodel achieves roughly a $10\times$ speedup compared to the layout. Note that in practice the VCO design may contain more complex circuitry which leads to longer runtime for a simulation run. The runtime for simulation with the Verilog-AMS module will not be substantially different. Thus the speedup will be more significant in that case.
Also note that the Verilog-AMS language along with the AMS simulator allow us to model and simulate the other blocks in the form of an HDL, which is a great advantage over the full transistor simulation.

\begin{table}[htbp]
\caption{Comparison of The Speed of The PLL Simulations with Different VCO Modules}
\label{TBL:speed}
\centering
\begin{tabular}{|c||c|c|c|}
\hline
{} & \bfseries  Layout & \bfseries Schematic & \bfseries Metamodel \\
\hline \hline
 Runtime & 80.5 s & 40.3 s & 8.7 s\\
 Normalized speed & $ 1 \times$ & $\sim 2 \times$ & $\sim 10 \times$\\
\hline
\end{tabular}
\end{table}

\section{PLL Optimization using Verilog-AMS-PAM}
\label{Sec:PLL_Optimization}

In this section, we demonstrate how Verilog-AMS-PAM can assist AMS system design space exploration using a PLL optimization example. The goal of this PLL optimization is to minimize the power dissipation ($P_D$) subject to the requirements for lock time ($T_L$), maximum frequency ($F_{T,max}$) and minimum frequency ($F_{T,min}$). The transistor sizes $W_P$ and $W_N$ of the LC VCO are chosen as the design variables, $x_1$ and $x_2$, to be optimized. Let $\mathbf{x} = \{ x_1, x_2 \}$. The optimization problem is formulated as:
\begin{equation}
\begin{array}{l l}
\text{minimize} & P_D(\mathbf{x}) \\
\text{subject to} &
\left\{
\begin{array}{l c l}
T_L(\mathbf{x}) & \leq & T_{L,min} = 400 \text{ ns} \\
F_{T,min}(\mathbf{x}) & \leq &  F_{T,min} = 2180 \text{ MHz} \\
F_{T,max}(\mathbf{x}) & \geq & F_{T,max} = 2300 \text{ MHz}.
\end{array}
\right.
\end{array}
\end{equation}

\begin{figure}[t]
\centering
\includegraphics[width=0.70\textwidth]{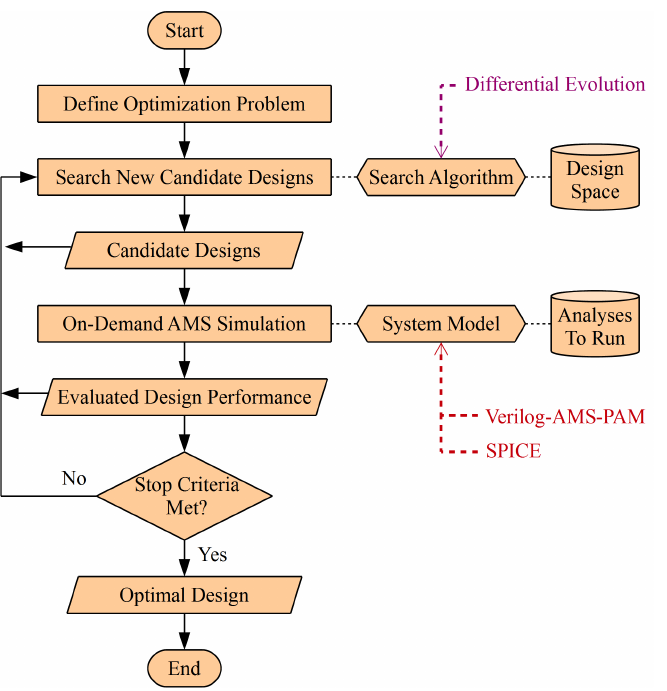}
\caption{A generic AMS system optimization flow employing a metaheuristic algorithm.}
\label{fig:optimization_flow}
\end{figure}

Metaheuristic algorithms are effective tools for solving analog optimization problems, such as in \cite{yu_2007,mcc_2009}. A generic AMS system optimization flow employing a metaheuristic algorithm is shown in Fig.~\ref{fig:optimization_flow}. It iteratively searches the design space to find the best design. The two key components in this flow are the search algorithm and the AMS system model. In this example, the differential evolution (DE) algorithm \cite{pri_2005} is selected as the search algorithm to demonstrate that the proposed framework is compatible to one of the popular metaheuristic algorithms. Other algorithms can of course be used as the proposed framework is algorithm-agnostic. 
The system model in the generic flow is for the evaluation of the objective and constraint functions. The speed of the evaluation greatly determines the speed of the optimization. The system model can consist of accurate but slow SPICE models or the efficient and layout-accurate proposed Verilog-AMS-PAMs. In this PLL optimization example, we compared the results of both.

\subsection{A Specific Optimization Algorithm - Differential Evolution}

 DE is a powerful direction-free algorithm that can handle nonlinear and non-differentiable objective functions. Thus, it is suitable for AMS optimization problems. In DE, the iterative process of finding the best design is described as an evolutionary development. The evolution starts with a population of candidate designs that are typically generated randomly but distributed over the entire design space. These candidate designs evolve from generation to generation following a specific scheme. This scheme consists of three steps: mutation, crossover, and selection. Every candidate design in every generation will experience this process. The DE algorithm customized for our PLL optimization is shown in Algorithm~\ref{ALG:DE}.

\begin{algorithm}[htbp]
\caption{A Specific Optimization Algorithm using \textbf{iVAMS 1.0} - Differential Evolution Algorithm.}
\label{ALG:DE}
\begin{algorithmic}[1]
\STATE Read DE parameters: scale factor $F$, crossover rate $CR$
\STATE Read current generation of $K$ candidate designs
\begin{displaymath}
\mathbf{X}_g = \{\mathbf{x}_{1,g}, \mathbf{x}_{2,g}, ..., \mathbf{x}_{K,g}\}
\end{displaymath}
\FOR{$i=1$ to $K$}
	\STATE Randomly choose three integers, $r1$, $r2$, $r3$ and ensure:
	\begin{displaymath}
	r1 \neq r2 \neq r3 \neq i \text{ and } \{r1, r2, r3\} \subset [1, K]
	\end{displaymath}
	\STATE Generate a mutant design $\mathbf{v} = \{ v_1, v_2, ..., v_D \}$:
	\begin{displaymath}
	\mathbf{v} = \mathbf{x}_{r1,g} + F \cdot (\mathbf{x}_{r2,g} - \mathbf{x}_{r3,g})
	\end{displaymath}
	\STATE Initialize a trial design $\mathbf{u} = \{ u_1, u_2, ..., u_D \}$:
	\begin{displaymath}
	\mathbf{u} \gets \mathbf{x}_{i,g}
	\end{displaymath}
	\FOR{$j=1$ to $D$}
		\IF {$randuni[0,1] \leq CR$ or $j = randint[1,D]$}
			\STATE $u_j \gets v_j$
		\ENDIF
	\ENDFOR
	\STATE Evaluate $P_D(\mathbf{u})$, $P_D(\mathbf{x}_{i,g})$, $T_L(\mathbf{x}_{i,g})$, $F_{T,max}(\mathbf{x}_{i,g})$, and $F_{T,min}(\mathbf{x}_{i,g})$ by running AMS simulation
	\STATE Constraint$_1 \gets T_L(\mathbf{x}_{i,g}) \leq  T_{L,min}$
	\STATE Constraint$_2 \gets F_{T,max}(\mathbf{x}_{i,g}) \geq F_{T,max}$
	\STATE Constraint$_3 \gets F_{T,min}(\mathbf{x}_{i,g}) \leq F_{T,min}$
	\STATE Initialize next generation of $K$ candidate designs:
	\begin{displaymath}
	\mathbf{X}_{g+1} \gets \mathbf{X}_{g}
	\end{displaymath}
	\IF {$P_D(\mathbf{u}) \leq P_D(\mathbf{x}_{i,g})$ and constraints are satisfied}
		\STATE $\mathbf{x}_{i,g+1} \gets \mathbf{u}$
	\ENDIF
\ENDFOR
\end{algorithmic}
\end{algorithm}

For a candidate design, mutation creates a mutant design by adding the difference, with a scale factor $F$, of two other designs that are randomly chosen from the current population to a third randomly chosen design. Crossover attempts to increase the population diversity by introducing a trial design. Based on a pre-determined crossover rate $CR$ and randomly generated numbers, the design variable values of the trial design are either from the candidate design or the mutant design. Selection evaluates the objective function for the candidate design and the trial design. The one with better performance is selected to be a member of the next generation and the inferior one is discarded. In the algorithm, the function $randuni[0,1]$ generates a uniformly distributed random number $\in [0,1]$ and $randint[1,D]$ produces a random integer uniformly distributed $\in [1,D]$.

\subsection{An Optimization Case Study - PLL Optimization}

\begin{figure}[htbp]
\centering
\includegraphics[width=0.55\textwidth]{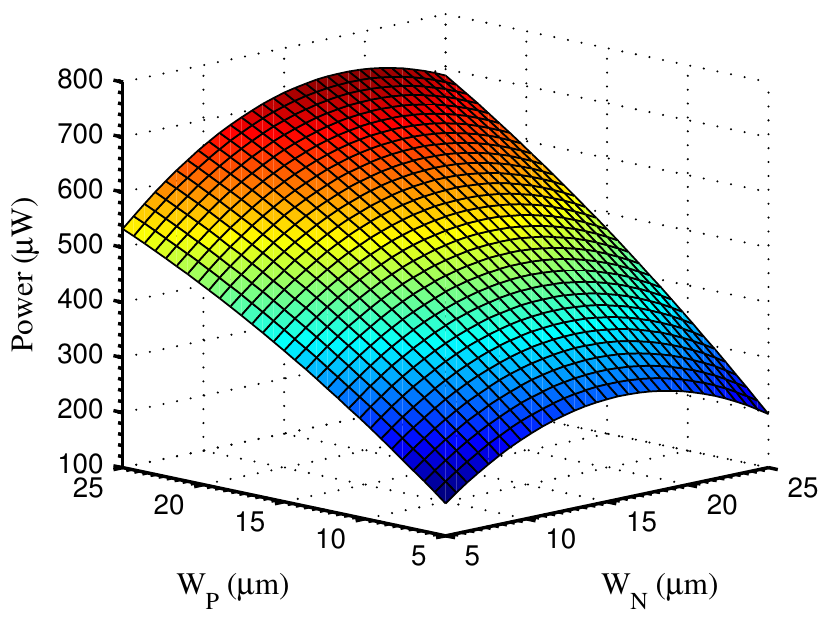}
\caption{A response surface of PLL power dissipation constructed from 900 simulation samples.}
\label{fig:surf}
\end{figure}

The ranges of the two design variables, $W_P$ and $W_N$, for the PLL optimization are both defined to be 5--25 $\mu$m. The response surface of the PLL power dissipation constructed from 900 simulation samples is shown in Fig.~\ref{fig:surf}. It reveals that this optimization task is not a particularly challenging problem. Still, it can serve to illustrate the effectiveness of Verilog-AMS-PAM in assisting AMS system optimization.

\begin{figure}[htbp]
\centering
\includegraphics[width=0.62\textwidth]{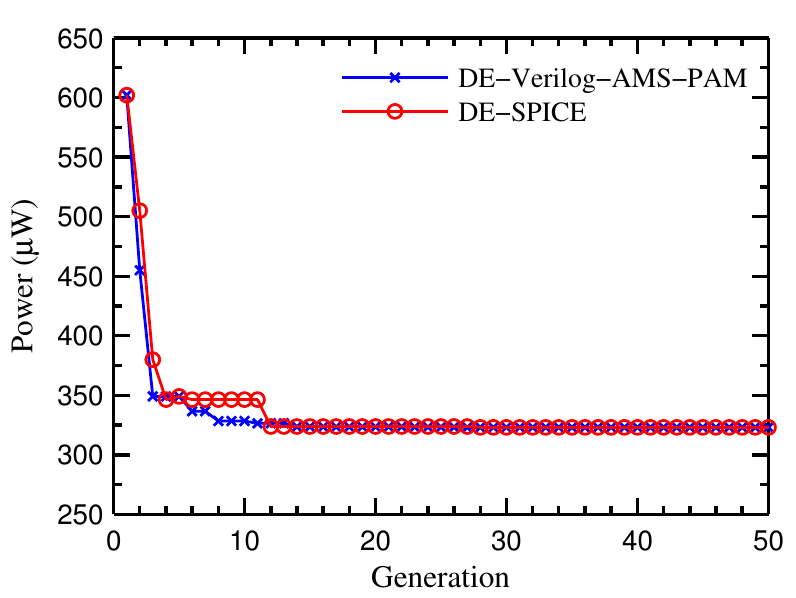}
\caption{Iterations of the DE optimization.}
\label{fig:iteration}
\end{figure}

The flow shown in Fig.~\ref{fig:optimization_flow} was used in the PLL optimization. Two optimization runs were performed. The first run, termed DE-Verilog-AMS-PAM, employed the DE algorithm shown in Fig.~\ref{ALG:DE} and used the LC-VCO Verilog-AMS-PAM in the PLL simulations for the evaluation of power, lock time, and tuning range. The second run, termed DE-SPICE, employed the same DE algorithm but used the LC-VCO layout netlist consisting of SPICE models in the PLL simulations. Both runs had the same classic DE setting \cite{pri_2005}: $F = 0.8$, $CR = 0.9$, $K = 20$, and $DE/rand/1/bin$. Both DE-Verilog-AMS-PAM and DE-SPICE ran until the 100th generation was reached. Fig.~\ref{fig:iteration} plots the best candidate designs in each generation. In both cases, the algorithm converged in 12 generations, so only 50, instead of 100, generations are plotted Fig.~\ref{fig:iteration}. Table~\ref{TBL:optim_speed} compares the runtime of the two optimization runs.

\begin{table}[t]
\caption{Comparison of the optimization runtime}
\label{TBL:optim_speed}
\centering
\begin{tabular}{|c|c|}
\hline
{\bfseries Optimization} & {\bfseries Runtime} (hours) \\
\hline
\hline
DE-Verilog-AMS-PAM & 5.06 \\
\hline
DE-SPICE & 45.55 \\
\hline
{ Runtime Reduction} & \multicolumn{1}{c|}{88.9 \%} \\
\hline
\end{tabular}
\end{table}

\begin{figure}[htbp]
\centering
\includegraphics[width=0.55\textwidth]{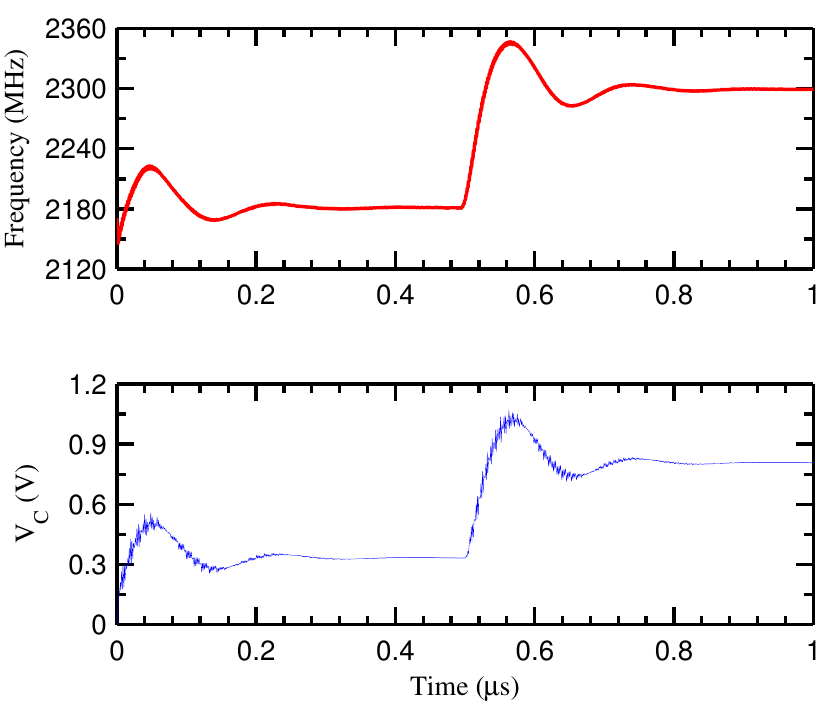}
\caption{Simulation showing that the PLL first locks to 2180 MHz and then relocks to 2300 MHz.}
\label{fig:relock}
\end{figure}

DE-Verilog-AMS-PAM and DE-SPICE found the same optimal design with 323 $\mu$W power dissipation. Table~\ref{TBL:optim} compares the baseline design and the optimal design. A 46~\% power savings was achieved through the DE algorithm. The lock time was also slightly improved by 1.5~\%. Fig.~\ref{fig:relock} shows that the PLL simulation with the the optimal design relocks from 2180 MHz to 2300 MHz. 

\begin{table}[htbp]
\caption{Comparison of the baseline and optimized designs}
\label{TBL:optim}
\centering
\begin{tabular}{|c||c|c|c|}
\hline
{} & \bfseries  Baseline & \bfseries Optimal & \bfseries Reduction \\
\hline \hline
{${{W_P}/{W_N}}$ } ($\mu$m$ / \mu$m) &  20~/~10  & 12.38~/~5 & -- \\
\hline
{Power} ($\mu$W) & 602 & 323 & 46~\% \\
\hline
{Lock time} (ns) & 335.4 & 330.4 & 1.5~\% \\
\hline
{Tuning Range} (MHz) & 2170--2304 & 2160--2394 & -- \\
\hline
\end{tabular}
\end{table}

\subsection{Computation Time Comparison Against Macromodels}

Section~\ref{Sec:PLL_Optimization} has shown the speed advantage of Verilog-AMS-PAM against SPICE models. For a typical AMS simulation flow, the block can be replaced with a macromodel with added parasitic effects. In order to make sense of the overall computation time with Verilog-AMS-PAM compared against that with traditional macromodels in the optimization of a large-scale mixed-signal system that requires long transient analyses, it is necessary to decompose their computation time. For each iteration, a flow with macromodels usually requires parameter re-extraction. Assuming the optimization requires $N_i$ iterations to reach the optimal design and the traditional macromodel is used, there are $N_i$ sets of circuit parameters to be extracted and $N_i$ transient analyses to be performed. Let $t_{ext}$ and $t_{sim}$ denote the computation time for each extraction and transient analysis. The total computation time for the macromodel based optimization is
\begin{equation}
\label{eq_tma}
t_{MA} = N_i \cdot t_{ext} + N_i \cdot t_{sim}.
\end{equation}
If the optimization is based on the Verilog-AMS-PAM, the total computation time can be expressed
\begin{equation}
\label{eq_tpom}
t_{PAM} = N_s \cdot t_{ext} + t_{gen} + N_i \cdot (t_{ini} + t_{sim}),
\end{equation}
where $N_s$ is the number of design samples used to construct Verilog-AMS-PAM, $t_{gen}$ is the time for generating Verilog-AMS-PAM, and $t_{ini}$ is the time that the AMS simulator takes to initialize and compile the Verilog-AMS-POM before running each transient analysis. Generally, $t_{gen}$ and $t_{ini}$ are very small. For example, in the presented case study, $t_{gen}$ is less than one minute, and $t_{ini}$ is less than one second. This is also true for more complex designs such as those in \cite{zheng_2012_isvlsi}. Therefore, they can be neglected from Equation~\ref{eq_tpom} which can then be reduced to
\begin{equation}
\label{eq_tpom2}
t_{PAM} = N_s \cdot t_{ext} + N_i \cdot t_{sim}.
\end{equation}
With Equations (\ref{eq_tma}) and (\ref{eq_tpom2}), the computational time difference between macromodel based and Verilog-AMS-PAM based optimization can be estimated as:
\begin{equation}
\label{eq_td}
t_{D} = t_{MA} - t_{PAM} = (N_i - N_s) \cdot t_{ext}.
\end{equation}
Assuming that the Verilog-AMS-PAM is constructed using 200 samples (or $N_s = 200$), that the optimization takes $N_i = 1200$ iterations, and that extracting circuit parameters for each design takes 60 seconds ($t_{ext} = 60 s$), the computation time reduction by using the Verilog-AMS-PAM based technique is $t_D \approx 16.7$ hours. Note that: 1) It has been assumed in this analysis that $N_i > N_s$, which is usually true since the response surface of the system is very likely more complex than a circuit block; 2) The time for re-doing layout or estimating parasitics, which would consume much more time for macromodel based flow, has not been included in the analysis. Equation (\ref{eq_td}) also reveals that Verilog-AMS-PAM is suitable for optimization algorithms that require large number of iterations but can converge to exceptional final designs.

\section{Conclusions and Future Research}
\label{Sec:Conclusions}

A method for fast mixed-signal system design space exploration using layout-accurate behavioral models has been proposed. A flow for creating circuit block behavioral models that accurately include physical design parasitics has been presented. Through a PLL case study, Verilog-AMS-PAM assisted AMS system verification and design space exploration have been demonstrated. The PLL optimization example demonstrates that the proposed Verilog-AMS-PAM is compatible with advanced optimization algorithms for AMS design optimization. For more complex designs or stringent operating conditions, more circuit parameters, such as mismatch characteristics, can be modeled as long as they can be extracted from circuit simulations.

Future research includes enhancing the capability of the Verilog-AMS-PAM based method of handling large and complex AMS systems. Such a system can be divided into multiple sub-systems each containing a number of circuit blocks. In such a case, creating behavioral models for various abstraction levels becomes necessary. At the block level, response surfaces can be sampled using SPICE simulations to ensure accuracy. This, however, is impractical because each SPICE simulation at this scale is very expensive. A potential solution is to create high-level Verilog-AMS-PAMs for each sub-system and to sample the response surfaces with Verilog-AMS simulation. This way the modeling speed can be greatly improved with minimum accuracy compromise thanks to the layout-level accuracy of Verilog-AMS-PAM.

The next version of iVAMS, \textbf{iVAMS 2.0} will include the integration of non-polynomial metamodels (such as machine learning based models) in Verilog-AMS
\cite{Mohanty_Book_2015_Mixed-Signal,oko_2012,Garitselov_GLSVLSI_2012}.

\section*{Acknowledgments}

A preliminary version of this research was presented in the following double-blind review conference: \cite{ZhengGLSVLSI2012}. \\
The authors would like to thank UNT graduate Dr. Geng Zheng for his help on this work.

\vspace{-0.2cm}
\bibliographystyle{bib/IEEEtran}


\vspace{-0.8cm}
\section*{Authors' Biographies}
\vspace{-0.2cm}

\begin{wrapfigure}{l}{0.95in}
\vspace{-0.5cm}
		\includegraphics[width=1.0in,keepaspectratio]{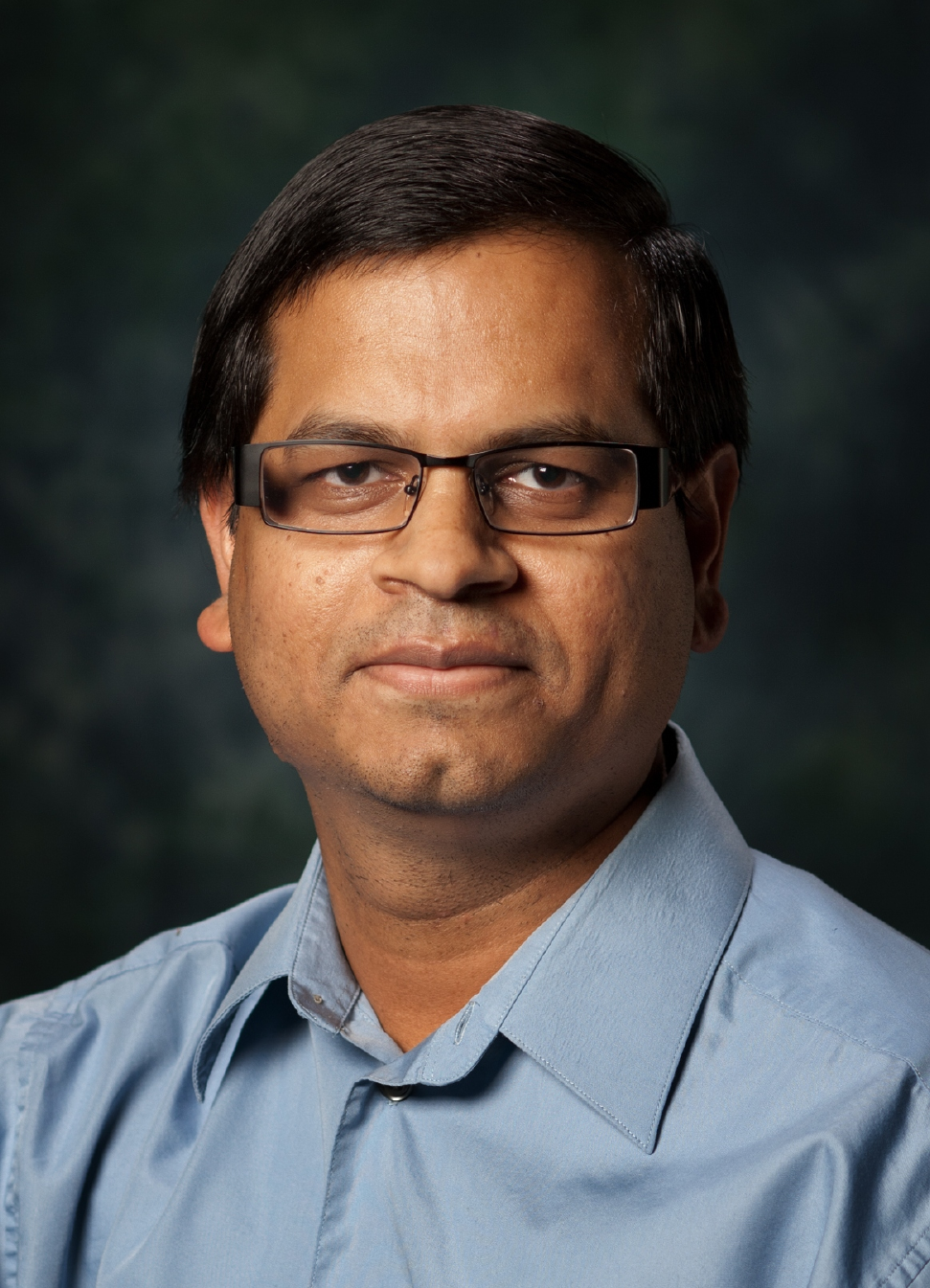}
\vspace{-0.9cm}
\end{wrapfigure}
\textbf{Saraju P. Mohanty} obtained a Bachelors degree with Honors in Electrical Engineering from the Orissa University of Agriculture and Technology (OUAT), Bhubaneswar, 1995. His Masters degree in Systems Science and Automation is from the  the Indian Institute of Science (IISc), Bangalore, in 1999. He obtained a Ph.D. in Computer Science and Engineering (CSE) in 2003, from the University of South Florida (USF), Tampa.
He is a Professor at the University of North Texas. His research is in ``Smart Electronic Systems'' which has been funded by National Science Foundations, Semiconductor Research Corporation, US Air Force, IUSSTF, and Mission Innovation Global Alliance.
He has authored 300 research articles, 4 books, and invented 4 US patents. His Google Scholar h-index is 31 and i10-index is 108. He has received 6 best paper awards and has delivered multiple keynote talks at various International Conferences. He received IEEE-CS-TCVLSI Distinguished Leadership Award in 2018 for services to the IEEE, and to the VLSI research community.
He has been recognized as a IEEE Distinguished Lecturer by the Consumer Electronics Society during 2017-2018. 
He was conferred the Glorious India Award in 2017 for his exemplary contributions to the discipline. He received Society for Technical Communication (STC) 2017 Award of Merit for his outstanding contributions to IEEE Consumer Electronics Magazine. 
He was the recipient of 2016 PROSE Award for best Textbook in Physical Sciences \& Mathematics category from the Association of American Publishers for his Mixed-Signal System Design book published by McGraw-Hill in 2015. 
He was conferred 2016-17 UNT Toulouse Scholars Award for sustained excellent scholarship and teaching achievements. 
He is the Editor-in-Chief of the IEEE Consumer Electronics Magazine. He served as the Chair of TC on VLSI, IEEE Computer Society during 2014-2018.

\vspace{0.2cm}
\begin{wrapfigure}{l}{0.95in}
	\vspace{-0.5cm}
	\includegraphics[width=1.0in,keepaspectratio]{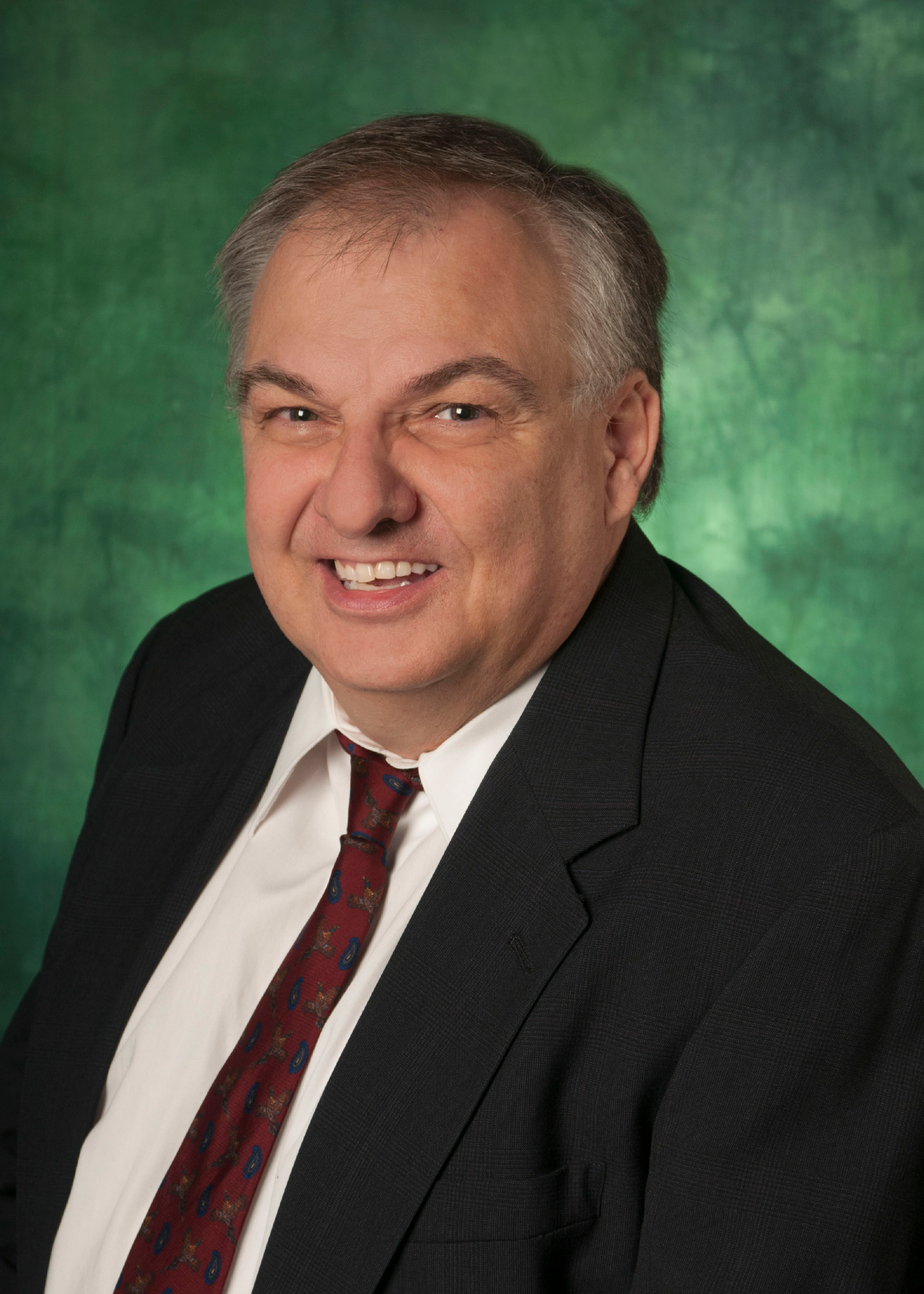}
	\vspace{-0.9cm}
\end{wrapfigure}
\noindent
\textbf{Elias Kougianos}  received a BSEE from the University of Patras, Greece in 1985 and an MSEE in 1987, an MS in Physics in 1988 and a Ph.D. in EE in 1997, all from Lousiana State University. 
From 1988 through 1997 he was with Texas Instruments, Inc., in Houston and Dallas, TX. Initially he concentrated on process integration of flash memories and later as a researcher in the areas of Technology CAD and VLSI CAD development. 
In 1997 he joined Avant! Corp. (now Synopsys) in Phoenix, AZ as a Senior Applications engineer and in 2001 he joined Cadence Design Systems, Inc., in Dallas, TX as a Senior Architect in Analog/Mixed-Signal Custom IC design. He has been at UNT since 2004. He is a Professor in the Department of Engineering Technology, at the University of North Texas (UNT), Denton, TX. His research interests are in the area of Analog/Mixed-Signal/RF IC design and simulation and in the development of VLSI architectures for multimedia applications. 
He is an author of over 120 peer-reviewed journal and conference publications. 

\end{document}